\documentstyle[11pt]{article}
\begin{document}
\renewcommand {\thepage} { }
\renewcommand {\thefootnote} {\fnsymbol{footnote}}
\setcounter {page} {0}
\setcounter {footnote} {0}
\vspace*{1cm}
\noindent FSUJ TPI QO-06/96
\begin{flushright}
April, 1996
\end{flushright}
\vspace{5mm}
\begin{center}{\Large \bf
Multi-mode density
matrices of light via amplitude and phase control
}\end{center}
\vspace{5mm}
\begin{center}
T. Opatrn\'{y}\footnote{Permanent address:
Palack\'{y} University, Faculty of Natural Sciences, Svobody 26,
77146 Olomouc, Czech Republic}, D.-G. Welsch \\
Friedrich-Schiller Universit\"at Jena,
Theoretisch-Physikalisches Institut \\
Max-Wien Platz 1, D-07743 Jena, Germany \\[1ex]
W.~Vogel \\
Universit\"at Rostock, Fachbereich Physik \\
Universit\"atsplatz~3, D-18051 Rostock, Germany 
\end{center}
%
\vspace{10mm}

\begin{center}{\bf Abstract}\end{center}
A new method is described for determining the quantum state
of correlated multimode radiation by interfering the modes and 
measuring the statistics of the
superimposed fields in four-port balanced homodyne detection.
The full information on the $N$-mode quantum state is obtained
by controlling both the relative amplitudes and the phases of 
the modes, which simplifies the reconstruction of 
density matrices to only $N$ $\!+$ $\!1$ Fourier transforms.
In particular, this method yields time-correlated multimode density matrices 
of optical pulses by superimposing the signal by a sequence of 
short local-oscillator pulses.

\vspace{1cm}

\vfill
\renewcommand {\thepage} {\arabic{page}}
\setcounter {page} {1}

Recently, increasing interest has been devoted to the problem of 
reconstruction of the quantum state of optical fields from measurable 
data. The feasibility of reconstruction of single-mode density matrices
was first demonstrated by Smithey et al. \cite{Smithey1} using optical 
homodyne tomography. In the experiments they determined both
the Wigner function and the density matrix in a field-strength basis. 
Later, the problem of direct reconstruction
of the density matrix in a field-strength basis and the photon-number
basis has been studied in Refs.~\cite{Kuehn1} and \cite{Ariano2},
respectively. All these methods are based on the fact that in 
balanced homodyne four-port detection the field-strength distributions
of a signal mode can be measured and knowledge of the field-strength 
distributions for all phases within a $\pi$-interval is equivalent 
to knowledge of the quantum state of the mode \cite{KVogel1}.
Whereas for determining the Wigner function a three-fold integral must
be calculated, the direct reconstruction of the density matrix elements 
can be accomplished with two integrals. The method has first been 
extended to two-mode quantum states \cite{Wignersymp} 
and later on the general situation 
for reconstructing $N$-mode quantum states 
from $N$-fold joint difference-count distributions 
has been investigated \cite{Kuehn2}. 
In these schemes the determinations of $N$-mode density matrices
and phase space functions, respectively, require
$2N$- and $3N$-fold integral transforms of the measured data. 

In the present paper we propose a new method for determining
$N$-mode density matrices ($N$ $\!>$ $\!1$) from the difference count 
distributions measurable in balanced homodyne four-port detection
with pre-superimposed signal fields.
Apart from the advantage that only two photodetectors are needed in that 
scheme, the reconstruction of the density matrices only requires 
$N\!+\!1$ rather than $2N$ integral transforms. 
The method does not only enable one to measure 
the (joint) density matrices of 
correlated $N$-mode fields very efficiently, but also yields 
$N$-mode density matrices of
optical pulses in terms of time-localized nonmonochromatic modes. 
The basic quantities in our scheme are the distributions of the 
sum field strengths of two or more modes undergoing both amplitude
and phase control. It is shown that these  
distributions yield the full information on the quantum state, provided
that they are recorded as functions of the relative field 
amplitudes of the modes and for phases within 
$\pi$-intervals.

We will illustrate the principle by dealing with the two-mode
case. An extension of the corresponding results to $N$ modes
($N$ $\!>$ $\!2$) is straightforward. 
Consider two typical experimental schemes as given in Fig.~1.
\begin{figure}[htb]
\unitlength9.5mm 
\begin{center}
\begin{picture}(16,6)

\put(0.5,3){\vector(1,0){1}}
\put(1.5,3){\vector(1,0){3}}
\put(4.5,3){\line(1,0){3}}

\put(5.5,4.5){\line(1,0){1}}

\put(2,1){\vector(0,1){1.5}}
\put(2,2.5){\line(0,1){2}}

\put(6,1){\vector(0,1){1.5}}
\put(6,2.5){\line(0,1){2}}

\put(7.5,2.5){\line(0,1){1}}


\put(0.985,2.025){\line(1,1){2}}
\put(1.025,1.985){\line(1,1){2}}

\put(4.985,2.025){\line(1,1){2}}
\put(5.025,1.985){\line(1,1){2}}

\put(7.5,3){\oval(1,1)[r]}
\put(6,4.5){\oval(1,1)[t]}

\put(5.5,0.3){\makebox(1,1){LO}}

\put(2.1,1.5){\makebox(0.5,0.5){$\hat a_{2}$}}
\put(0.5,3.1){\makebox(0.5,0.5){$\hat a_{1}$}}


\put(3.5,3.1){\makebox(1.5,0.5){$\hat b$}}

\put(0.5,5){\makebox(1,1){{\Large $(a)$}}}


\put(9,3){\vector(1,0){2}}
\put(11,3){\line(1,0){3}}

\put(12.5,1){\vector(0,1){1.5}}
\put(12.5,2.5){\line(0,1){2}}

\put(12,4.5){\line(1,0){1}}
\put(14,2.5){\line(0,1){1}}

\put(14,3){\oval(1,1)[r]}
\put(12.5,4.5){\oval(1,1)[t]}

\put(11.535,2.025){\line(1,1){2}}
\put(11.565,1.985){\line(1,1){2}}

\put(12,0.3){\makebox(1,1){LO}}

\put(10.75,3){\oval(2.5,0.4)[t]}

\put(12.72,1.5){\oval(1.6,0.15)[r]}
\put(12.72,1.775){\oval(0.4,0.4)[bl]}
\put(12.72,1.225){\oval(0.4,0.4)[tl]}

\put(12.72,2){\oval(1,0.15)[r]}
\put(12.72,2.275){\oval(0.4,0.4)[bl]}
\put(12.72,1.725){\oval(0.4,0.4)[tl]}

\put(13.8,1.8){\makebox(0.5,0.5){$f_{1}(t)$}}
\put(13.8,1.3){\makebox(0.5,0.5){$f_{2}(t)$}}

\put(9.5,3.3){\makebox(2,0.5){signal}}

\put(9.5,5){\makebox(1,1){{\Large $(b)$}}}

\end{picture}
\end{center}
\caption{
Two possible schemes for reconstructing two-mode density matrices.
\protect\\
$(a)$ Two modes ($\hat a_1$ and $\hat a_2$) are mixed by
a beam splitter and one of the interfering output modes ($\hat b$) 
is used as signal mode in balanced homodyning in order to measure
the sum field strengths of the two modes (LO, strong local oscillator).
\protect\\ 
$(b)$ A signal pulse and a sequence of two short (strong)
local-oscillator pulses with envelopes $f_1(t)$ and $f_2(t)$ 
are superimposed in balanced homodyne detection. 
}
\label{fig1}
\end{figure}
First, let us assume that two 
(correlated) modes can be used separately as input 
fields in an interferometer, such as a beam splitter. 
The superimposed light in one output channel 
of the interferometer is used as a signal mode in 
balanced four-port homodyne detection [Fig.~\ref{fig1}$(a)$]. 
Second, two-mode density matrices of optical pulses can 
be measured [Fig.~\ref{fig1}$(b)$] by superimposing 
a signal pulse with a sequence of two short (strong) 
local oscillator (LO) pulses and recording the (time-integrated)
difference-count distributions \cite{Opatrny1}.
The time-localized (nonmonochromatic) modes are defined by the
short local-oscillator pulse envelopes $f_1(t)$ and $f_2(t)$,
and the measured difference-count distributions yield the 
distributions of the sum field strengths of the two modes
\cite{Munroe1}. 
In both schemes, the relative amplitude of the two modes 
and their phases are controlled.

Consider a two-mode optical field with photon destruction 
operators $\hat a_{k}$ ($k$ $\!=$ $\!1,2$) and introduce 
the scaled field-strength operators of the modes as 
\begin{equation}
\hat F_{k}(\varphi_k ) = |F| \left( \hat a_{k} e^{-i\varphi_k}
+ \hat a_{k}^{\dagger} e^{i \varphi_k} \right),
\label{0}
\end{equation}
and $\hat F_{k}(\varphi_k)|{\cal F}_k,\varphi_k\rangle  = {\cal F}_k
|{\cal F}_k,\varphi_k\rangle $. The mode amplitude 
$|F|$ specifies the kinds of fields (electric, magnetic) under study, 
to compare with experiments it can be related to 
the shot noise \cite{Vogel1}. 
In the field strength basis the density matrix is given 
by \cite{Kuehn2,Vogel1}
\begin{eqnarray}
\lefteqn{
\big\langle{\cal F}_1-{\cal F}_1',{\cal F}_2-{\cal F}_2',
\varphi_1,\varphi_2|\,\hat\varrho\,|
{\cal F}_1+{\cal F}_1',{\cal F}_2+{\cal F}_2',
\varphi_1,\varphi_2\big\rangle
}
\nonumber \\ && \hspace{2ex}
= \left(\frac{1}{2\pi}\right)^2
\int {\rm d}y_1 \int {\rm d}y_2 \,
e^{-i(y_1{\cal F}_1+y_2{\cal F}_2)}
\, \Psi(z_1,z_2,\psi_1,\psi_2),
\label{3}
\end{eqnarray}
where
\begin{equation}
z_k \equiv z_k(y_k,{\cal F}_k')
=\sqrt{{y_k}^2+{\cal F}_k'^2/|F|^4},
\label{1a}
\end{equation}
\begin{equation}
\psi_k \equiv \psi_k(y_k,{\cal F}_k')
= \varphi_k - {\rm arccot}[y_k |F|^2/{\cal F}_k']
\label{1b}
\end{equation}
(${\cal F}_k'$ $\!>$ $\!0$). The characteristic function 
$\Psi$ in Eq.~(\ref{3}) is closely related to the average 
of the two-mode coherent displacement operator and can be 
written as
\begin{equation}
\Psi(z_1,z_2,\psi_1,\psi_2)
=\left\langle\exp\!\left[iz_1\hat{F}_1(\psi_1)
+iz_2\hat{F}_2(\psi_2)\right]\right\rangle\!.
\label{1c}
\end{equation}
We see that $\Psi$ as a function of the $z_k$ (for given $\psi_k$) 
is nothing but the characteristic function of the joint field-strength
distribution
\begin{eqnarray}
\label{2}
\lefteqn{
p_{\rm j}({\cal F}_{1},{\cal F}_{2},\psi_{1},\psi _{2})
\equiv \langle {\cal F}_{1},{\cal F}_{2},\psi_1,\psi_2 
| \, \hat \varrho \, |
{\cal F}_{1},{\cal F}_{2},\psi_1,\psi_2 \rangle 
}
\nonumber \\ && \hspace{2ex}
= \left( \frac{1}{2\pi}\right) ^{2} 
\int {\rm d} z_{1}  \int {\rm d} z_{2} \,
e^{-i(z_{1}{\cal F}_{1} + z_{2}{\cal F}_{2})}
\,\Psi(z_1,z_2,\psi_1,\psi_2).
\end{eqnarray}
Thus, when the joint probability distributions of the two field 
strengths are measured for all values of the phases $\psi _{k}$
within $\pi$ intervals, two-fold Fourier transforms yield
their characteristic functions and two more Fourier integrals must
be performed to obtain the density matrix in Eq.~(\ref{3}).

Let us now suppose that in place of the above mentioned
joint field-strength distributions the distributions  
$p_{\rm s}({\cal F},\alpha,\psi_1,\psi_2)$ of the sum field 
strength
\begin{equation}
\hat{F} = \hat{F}_1(\psi_1) \cos\alpha 
+ \hat{F}_2(\psi_2) \sin\alpha
\label{2a}
\end{equation}
are measured, where $\alpha$ $\! \in \!$ $\!(0, \frac{1}{2}\pi)$ controls the
relative field amplitude.
The probability distributions $p_{\rm s}({\cal F},\alpha,\psi_1,\psi_2)$ 
and $p_{\rm j}({\cal F}_1,{\cal F}_2,\psi_1,\psi_2)$
are related to each other as
\begin{equation}
\label{3a}
p_{\rm s}({\cal F},\alpha , \psi _{1} , \psi _{2})
= \int \mbox{d}{\cal F}_{1} \int \mbox{d}{\cal F}_{2} \,
p_{\rm j}({\cal F}_{1},{\cal F}_{2},\psi _{1} , \psi _{2})
\,\delta ({\cal F} \!-\! {\cal F}_{1}\cos \alpha 
\!-\! {\cal F}_{2}\sin \alpha ), 
\end{equation}
which together with Eq.~(\ref{2}) implies that
\begin{equation}
\label{4}
p_{\rm s}({\cal F},\alpha,\psi_{1},\psi _{2})
= \frac{1}{2\pi} \int {\rm d} z \,
e^{-i z {\cal F}}
\,\Psi(z\cos\alpha,z\sin\alpha,\psi_1,\psi_2),
\end{equation}
and hence
\begin{equation}
\label{4b}
\Psi(z_1,z_2,\psi_1,\psi_2)
= \int {\rm d}{\cal F} \, e^{iz{\cal F}}
\,p_{\rm s}({\cal F},\alpha,\psi_1,\psi_2),
\end{equation}
where
\begin{equation}
\label{4b1}
z \equiv z(z_1,z_2) = \sqrt{z_1^2+z_2^2},
\end{equation}
\begin{equation}
\label{4c}
\alpha \equiv \alpha(z_1,z_2) = \arctan(z_2/z_1).
\end{equation}
This means that from measurements of the superimposed light (\ref{2a})
for different $\alpha$ we can  ``tomographically'' reconstruct the joint
characteristic function (\ref{4b}).
Combining Eqs.~(\ref{3}) and (\ref{4b}) yields
\begin{eqnarray}
\lefteqn{
\big\langle{\cal F}_1-{\cal F}_1',{\cal F}_2-{\cal F}_2',
\varphi_1,\varphi_2|\,\hat\varrho\,|
{\cal F}_1+{\cal F}_1',{\cal F}_2+{\cal F}_2',
\varphi_1,\varphi_2\big\rangle
}
\nonumber \\ && \hspace{2ex}
= \left(\frac{1}{2\pi}\right)^2
\int {\rm d}y_1 \int {\rm d}y_2 \,
e^{-i(y_1{\cal F}_1+y_2{\cal F}_2)} 
\int {\rm d}{\cal F} \,
e^{iy{\cal F}}\, p_{\rm s}({\cal F},\beta,\psi_1,\psi_2),
\label{4a}
\end{eqnarray}
where [according to Eqs.~(\ref{1a}), (\ref{4b1}), and (\ref{4c})]
\begin{equation}
\label{4d}
y \equiv y(y_1,y_2,{\cal F}_1',{\cal F}_2') = 
\sqrt{y_1^2+y_2^2+({\cal F}_1'^2+{\cal F}_2'^2)/|F|^4}\,,
\end{equation}
\begin{equation}
\label{4e}
\beta \equiv \beta(y_1,y_2,{\cal F}_1',{\cal F}_2')
= \arctan\sqrt{\frac{y_2^2|F|^4+{\cal F}_2'^2}{y_1^2|F|^4+{\cal F}_1'^2}}\,,
\end{equation}
and $\psi_k$ is given in Eq.~(\ref{1b}).
Equation~(\ref{4a}) explicitly shows the feasibility of reconstructing 
the two-mode density matrix from the measured sum 
field-strength distributions 
[$\psi_k$ $\! \in \!$ $\!(\varphi_k$ $\! -$ $\!\pi, \varphi_k)$,
$\beta$ $\! \in \!$ $\!(0, \frac{1}{2}\pi)$].

Let us briefly comment on the effect of nonperfect detection. 
When the photodetectors have efficiencies $\eta$ that are less than
unity, the characteristic function of the (sum) field strength distribution
in Eq.~(\ref{4a}) is replaced according to \cite{Vo-Gr}
\begin{equation}
\int {\rm d}{\cal F} \,
e^{iy{\cal F}}\, p_{\rm s}({\cal F},\beta,\psi_1,\psi_2)
\to 
e^{y^2|F|^2(1-\eta)/(2\eta)}
\!\int \! {\rm d}{\cal F} \,
e^{iy{\cal F}}  
\, p_{\rm s}({\cal F},\beta,\psi_1,\psi_2;\eta),
\label{4ee}
\end{equation}
where $p_{\rm s}({\cal F},\beta,\psi_1,\psi_2;\eta)$ denotes the sum field 
distribution measured with a detector of quantum efficiency $\eta$.
Consequently, Eq.~(\ref{4a}) is generalized as
\begin{eqnarray}
\lefteqn{
\big\langle{\cal F}_1-{\cal F}_1',{\cal F}_2-{\cal F}_2',
\varphi_1,\varphi_2|\,\hat\varrho\,|
{\cal F}_1+{\cal F}_1',{\cal F}_2+{\cal F}_2',
\varphi_1,\varphi_2\big\rangle
}
\nonumber \\ && 
= \!\!\left(\frac{1}{2\pi}\right)^2
\!\int \! {\rm d}y_1 \! \int \! {\rm d}y_2 \,
e^{-i(y_1{\cal F}_1+y_2{\cal F}_2)}
\,e^{y^2|F|^2(1-\eta)/(2\eta)}
\!\int \! {\rm d}{\cal F} \,
e^{iy{\cal F}}  
\, p_{\rm s}({\cal F},\beta,\psi_1,\psi_2;\eta).
\nonumber \\ &&
\label{6}
\end{eqnarray}
Processing data from a real experiment
requires a careful handling of the noise to avoid troubles arising from
the exponential of $y^{2}$.

Let us return to the measurement schemes proposed above. Equation~(\ref{6})
yields the general relation of the measured distributions to the 
corresponding two-mode density matrices. In the first scheme 
[Fig.~\ref{fig1}$(a)$] the difference
count distributions directly yield the desired  
distributions $p({\cal F},\beta,\psi_1,\psi_2;\eta)$ of the sum field
of the two signal modes under study.
The parameters $\psi_1$, $\psi_2$, and $\beta$ are related to 
the input--output relations of
the interferometer. Measurement of the difference-count distributions
on a sufficiently dense grid of points ($\psi_1,\psi_2,\beta$),
with $\psi_k$ $\! \in \!$ $\!(\varphi_k$ $\!-$ $\!\pi, \varphi_k)$ 
and $\beta$ $\! \in \!$ $\!(0, \frac{1}{2}\pi)$, is then equivalent
to measurement of the quantum state of the two-mode field.
In the second scheme [Fig.~\ref{fig1}$(b)$], two-mode density matrices of 
optical pulses can be measured by superimposing 
the signal pulse with two short (strong) 
local oscillator (LO) pulses of envelopes $f_1(t)$ and $f_2(t)$.
The desired information on the two-mode quantum state is obtained by
recording the (time-integrated)
difference-count distributions and performing the amplitude
and phase control by controlling the relative strength and the positions
of the LO pulses, for more details see \cite{Opatrny1}.
Note that this method is not only of interest for studying pulses, it also
yields insight in the time-dependent correlation properties of stationary
fields at the level of the full quantum statistical information.

In certain cases of pulse measurements, when the signal pulse and the LO 
pulses are produced by different sources, it may become impossible to  
achieve complete phase control. However, the difference of the LO phases, 
$\Delta \psi$ $\! =$
$\!\psi _{2}$ $\! -$ $\! \psi _{1}$, can be controlled
and we  measure the phase averaged probability distribution,
\begin{equation}
\overline{p_{s}({\cal F},\beta , \Delta \psi)}\equiv
(2\pi)^{-1}\int \mbox{d} \psi _{1} p_{s}({\cal F},\beta , \psi _{1},
\psi _{1} + \Delta \psi).
\end{equation}
Concequently, we can reconstruct the phase-averaged (with respect to 
$\psi_1$) version of the density matrix in Eq.~(\ref{6}).
Although losing some information, these averaged density matrices still
contain interesting informations on the temporal correlations of the
signal field.

It is straightforward to extend the method to measurements
of $N$-mode density matrices ($N$ $\!>$ $\!2$).
Again, we can measure the weighted sums of the field strengths in 
a homodyne setup with two detectors. The pre-superimposed signal 
can again be obtained using appropriate interferometric methods. 
In particular, in the scheme in Fig.~\ref{fig1}$(b)$ a train of 
$N$ (strong) LO pulses must be used. The sum field strengths 
are now  parameterized by $N$ $\!-$ $\!1$ relative intensities 
of the $N$ modes under consideration, and the characteristic 
functions of their probability densities are related to the 
characteristic functions of the joint probability densities 
by generalizing Eq.~(\ref{4}). Then the reconstruction of the 
$N$-mode density matrix can be accomplished with $N$ $\!+$ $\!1$ 
Fourier integrals  -- one integral for obtaining the characteristic 
function of the density matrix from the measured data 
plus one integral per mode in order to determine the density 
matrix from its characteristic function.

In conclusion, we have shown that the reconstruction of multimode density
matrices of light can be performed by measuring amplitude and phase 
controlled distributions of the sum fields. This simplifies the
reconstruction of multimode density matrices by reducing both the number of 
photodetectors needed and the number of
the desired integral transforms. Applications of the method for
determining the quantum states of correlated field modes 
and of time-correlated multimode quantum states are discussed.
The latter allows to perform time-dependent correlation measurements 
yielding the full quantum statistical information.

\vspace{3ex}

We thank R. Sauerbrey for useful discussions.
This work was supported by the Deutsche Forschungsgemeinschaft.

\end{document}